\documentclass[%
 superscriptaddress,
 groupedaddress,
 aip,
 apm,%
 amsmath,amssymb,
 reprint,%
]{revtex4-2}
\usepackage{xcolor}
\usepackage{hyperref}
\usepackage{graphicx}
\usepackage{dcolumn}
\usepackage{bm}
\usepackage{ragged2e} 
\usepackage{tabularx}
\usepackage{setspace}

\newcolumntype{L}{>{\RaggedRight\arraybackslash}X}

\begin{document}

\preprint{AIP/123-QED}

\title{\textit{Ab initio} thermal conductivity of Ge$_x$Sn$_{1-x}$O$_2$ alloys}

\author{Xiao Zhang}
 \affiliation{Department of Materials Science and Engineering, University of Michigan, Ann Arbor, 48109, USA}
\author{Emmanouil Kioupakis}%
 \email{kioup@umich.edu}
\affiliation{Department of Materials Science and Engineering, University of Michigan, Ann Arbor, 48109, USA}

\date{\today}

\begin{abstract}
Rutile GeO$_2$ is an emerging ultra-wide band gap semiconductor (UWBG) that has demonstrated excellent potential for applications in power electronic devices. Alloys of rutile SnO$_2$, a well-established UWBG semiconducting oxide, with GeO$_2$ are promising for tuning the material properties for applications. The thermal conductivity, in particular, is a key property which is significantly impacted by alloy disorder, but which is also essential in assessing the operation and degradation of materials in high-power electronic applications. Here, we present first-principles calculations of the thermal conductivity of rutile GeO$_2$, SnO$_2$, and their alloys, and quantify the effects of scattering by alloy disorder, temperature, and isotope mass distribution. We show that the relatively high thermal conductivity of the binary compounds is reduced by alloying, grain boundaries, and isotope disorder. However, we also find that the room-temperature thermal conductivity of the alloys is still comparable to or surpasses the values for $\beta$-Ga$_2$O$_3$, an established UWBG semiconducting oxide. Our findings provide a roadmap for the codesign of the thermal properties of rutile Ge$_x$Sn$_{1–x}$O$_2$ alloys for electronic device applications.
\end{abstract}

\maketitle

\section{Introduction}

SnO$_2$ is a well-established ultra-wide band gap (UWBG) semiconductor. With a band gap of approximately 3.7 eV and excellent carrier transport capability\cite{wang2024,jiang2018,10.1063/1.1660648,10.1063/1.5018983}, the material has been widely studied for many applications such as sensors\cite{GOPEL19951,DAS2014112} and transparent conductors\cite{PhysRevLett.88.095501,BENSHALOM199320,10.1116/1.1894421,10.1063/1.3671162,PhysRevB.92.235201}. Additionally, GeO$_2$, which under ambient conditions adopts the same rutile crystal structure as SnO$_2$, is an emerging UWBG semiconductor that in recent years has gained significant attention for power
electronics applications\cite{LABED2025,deng2021,RATHORE2022413466}. 
It has been predicted to have a potential ambipolar dopability\cite{chae2019,chae2021}, high carrier mobilities for both electrons and holes\cite{bushick2020}, and a high breakdown field due to its ultra-wide electronic band gap\cite{chae2021}. Experimental realizations of high-quality r-GeO$_2$ have been demonstrated via several different routes\cite{Takane_2021h, takane2021e,chae2022,deng2021,abed2024,rahaman2024,GOLBASI2025178591}. Possible doping techniques have been explored\cite{Essajai_2022,niedermeier2020} and efficient $n$-type doping of bulk crystals with Sb$^{5+}$ has been successfully demonstrated above $10^{20}$ cm$^{-3}$.\cite{https://doi.org/10.1002/pssb.202400326}

Alloying GeO$_2$ with SnO$_2$ introduces the possibility of tuning the material properties\cite{takane2022,kluth2024,nagashima2022}, providing an opportunity to develop alloys optimized for specific application needs. The experimental growth of Ge$_x$Sn$_{1-x}$O$_2$ alloys has been reported using pulsed laser deposition (PLD)\cite{nagashima2022,kluth2024}, chemical vapor deposition (CVD)\cite{takane2022}, and molecular beam epitaxy (MBE)\cite{doi:10.1021/acs.nanolett.4c05043} methods across a range of alloy compositions. Interestingly, it has been demonstrated that the alloys exhibit carrier mobilities that are not very sensitive to composition up to x $\sim$ 0.57\cite{takane2022,nagashima2022}. 
This insensitivity indicates that Ge$_x$Sn$_{1-x}$O$_2$ alloys can maintain favorable electronic transport properties even with significant changes in composition, which is advantageous for device design and manufacturability.

While the electronic properties of Ge$_x$Sn$_{1-x}$O$_2$ alloys are promising, their thermal conductivities must also be carefully evaluated to ensure their effectiveness in high-power applications. 
In power electronics, devices must dissipate substantial amounts of heat to maintain stability and prevent degradation over time. 
A high thermal conductivity enables efficient heat dissipation, which is particularly important for materials expected to function in compact or high-power-density configurations. While both SnO$_2$ and GeO$_2$ are known for their high thermal conductivities\cite{galazka2020transparent,chae2020}, the presence of alloy disorder often leads to significant phonon scatterings and thus a corresponding reductions in the thermal conductivity, a phenomenon observed in many other alloy systems\cite{fit1,fit2,Huang_2022,liu2004,ma16113945}. 
Therefore, to assess the potential of Ge$_x$Sn$_{1-x}$O$_2$ alloys for power electronic applications, it is essential to quantify the effect of alloy disorders on their thermal conductivities.

In this study, we employ first-principles calculations to consistently investigate the lattice thermal conductivity of GeO$_2$, SnO$_2$, and Ge$_x$Sn$_{1-x}$O$_2$ alloys as a function of composition, temperature, and isotope disorder. Our findings show that alloying significantly reduces the thermal conductivity compared to the binary compounds. 
However, even the lowest conductivity value remains similar to $\beta$-Ga$_2$O$_3$, another UWBG semiconducting oxide. Furthermore, by analyzing the impact of phonon mean-free path (MFP) and isotope scattering on thermal properties, we provide insights into the fundamental factors that limit the thermal conductivity of these alloys.

\section{Computational methods\label{sec:comput}}

First-principles calculations are performed with density functional theory\cite{PhysRev.140.A1133,PhysRev.136.B864} and related approaches. 
Structural relaxations are performed with the Quantum Espresso (QE)\cite{giannozzi2009quantum,giannozzi2017advanced} package, using norm-conserving pseudopotentials\cite{FUCHS199967} and the local density approximation for the exchange-correlation functional.\cite{PhysRevB.23.5048,PhysRevLett.45.566} 
For Ge and Sn, the 3d and 4d electrons are included in the valence, respectively, to ensure accurate phonon frequencies. Plane-wave energy cutoffs of 140 Ry are used to ensure the convergence of the total energy within 1 meV/atom. 
The unit cells are relaxed until the forces on the atoms are less than $5\times10^{-5}$ Ry/Bohr, and the total stresses on the unit cells are less than $5\times10^{-6}$ Ry/Bohr$^3$. Density functional perturbation theory\cite{RevModPhys.73.515} is used to evaluate vibrational properties within the harmonic approximation to obtain the second-order (harmonic) force constants, implemented in the QE package. Phonon frequencies are evaluated on a Brillouin zone (BZ) sampling grid of $4\times4\times6$, therefore producing the second-order force constants corresponding to a $4\times4\times6$ supercell.

To evaluate the lattice thermal conductivity, third-order anharmonic force constants are required, which are obtained through finite-difference approach with atomic displacements using $4\times4\times4$ supercells. 
Interactions up to third-nearest neighbors are considered. The structures, force constants, and Born effective charges are supplied to the almaBTE\cite{CARRETE2017351} code to solve the phonon Boltzmann transport equation in order to evaluate the thermal conductivity of GeO$_2$ and SnO$_2$. 
To evaluate thermal conductivities of the alloys, the virtual crystal approximation is used which takes the arithmetic averages\cite{CARRETE2017351}: $\phi_{VC}=\sum_{i}x_i\phi_i$, where $\phi_{VC}, \phi_i$, and $i$ are the physical properties (lattice constants, atomic coordinates, force constants) of the virtual crystal, the end compound, and the mole fraction of the end compound, respectively. 
Alloy and isotope mass disorders are considered throughout the calculations in this article. 
The average atomic mass with the natural isotope distribution\cite{berglund2011isotopic} of Ge (72.64 amu) and Sn (118.71 amu) are used. 
The effect of isotope mass disorder is examined separately for the binary compounds using ShengBTE\cite{ShengBTE_2014}. 
The isotopes considered according to natural isotope distribution\cite{berglund2011isotopic,ShengBTE_2014} are for Ge: Ge$^{70}$ (20.5\%), Ge$^{72}$ (27.4\%), Ge$^{73}$ (7.8\%), Ge$^{74}$ (36.5\%), and Ge$^{76}$ (7.8\%). For Sn: Sn$^{112}$ (0.97\%), Sn$^{114}$ (0.65\%), Sn$^{115}$ (0.36\%), Sn$^{116}$ (14.7\%), Sn$^{117}$ (7.7\%), Sn$^{118}$ (24.3\%), Sn$^{119}$ (8.6\%), Sn$^{120}$ (32.4\%), Sn$^{122}$ (4.6\%), and Sn$^{124}$ (5.6\%). 
A BZ sampling grid of $16\times16\times24$ is used consistently to evaluate the thermal conductivities of SnO$_2$, GeO$_2$, and the alloys.

\section{Phonon properties\label{sec:phonon}}

\begin{figure}
    \centering
    \includegraphics[width=0.49\columnwidth]{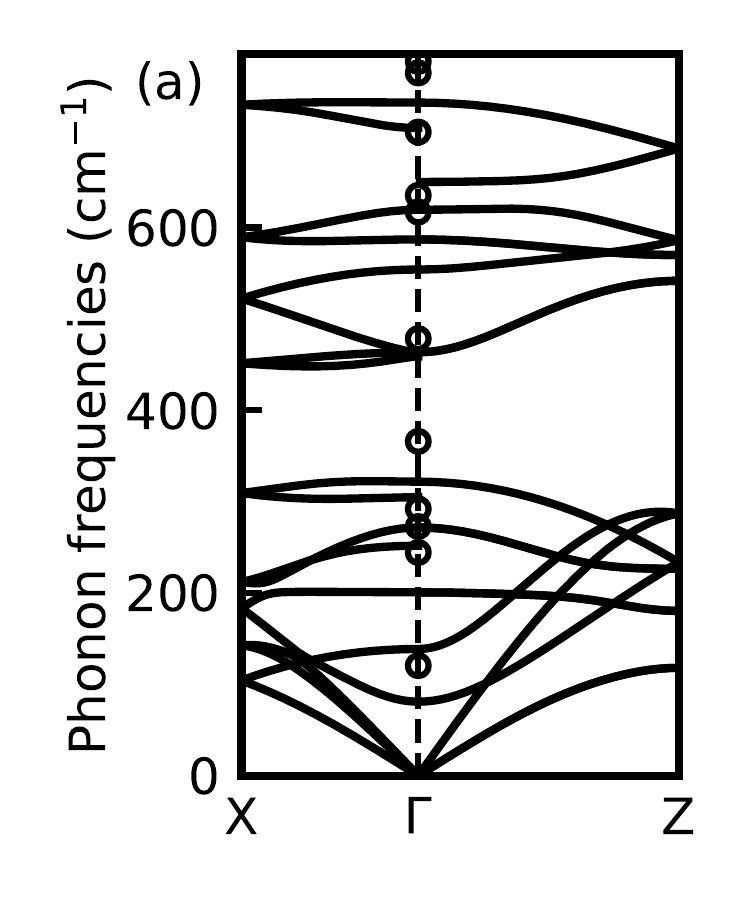}
    \includegraphics[width=0.49\columnwidth]{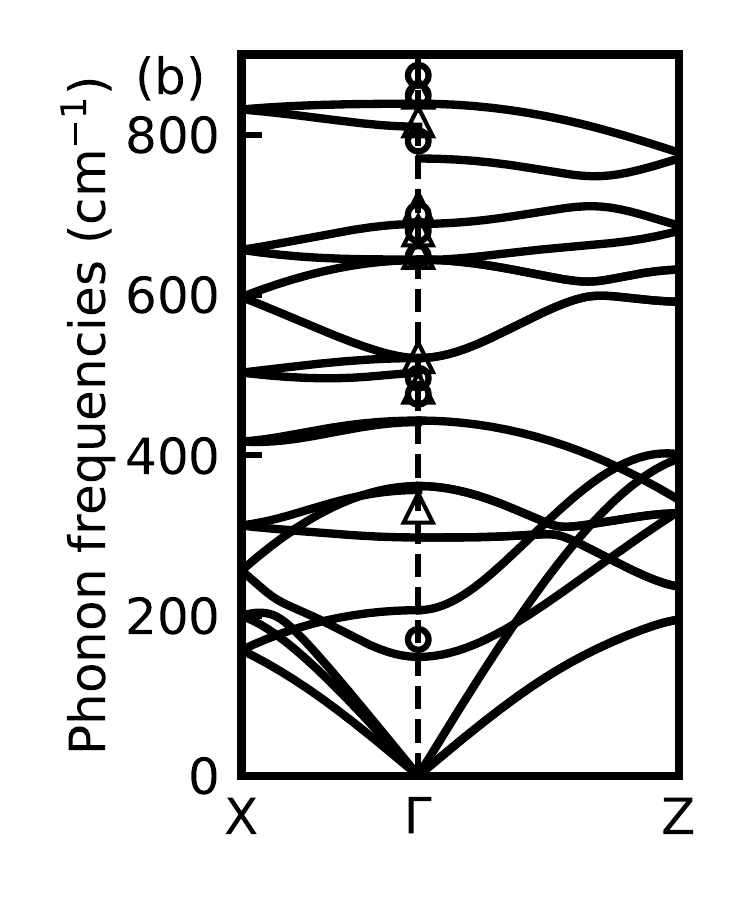}
    \caption{Phonon dispersion of (a) SnO$_2$ and (b) GeO$_2$ along the $X-\Gamma-R$ high-symmetry path of the first Brillouin zone. Our results are in good agreement with theoretical calculations from Ref.\onlinecite{wang2024} for SnO$_2$ and Ref.\onlinecite{chae2020} for GeO$_2$. Good agreement is also seen compared to experimental measurements for SnO$_2$ (Ref.\onlinecite{PhysRevB.86.134302}) and GeO$_2$. (Circles: Ref.\onlinecite{kahan1971polarized}, Triangles: Ref.\onlinecite{kaindl2012quantum})}
    \label{fig:phband}
\end{figure}

The calculated phonon dispersion relations for rutile SnO2 and GeO2 along the $\perp c$ and $\parallel c$ directions are shown in Figure ~\ref{fig:phband}. The calculated phonon frequencies are in good agreement with previous calculations\cite{wang2024,chae2020} and experimental measurements\cite{kahan1971polarized,kaindl2012quantum,PhysRevB.86.134302}. 
Overall, the phonon frequencies of GeO$_2$ are higher than SnO$_2$, which is attributed to the lighter atomic mass of Ge. 
We further evaluate the sound velocity based on the slope of the acoustic phonon branches near the $\Gamma$-point, as shown in Table \ref{tab:v_sound}. 
Our calculated sound velocities agree well with the reported values in the literature\cite{bushick2020,wang1973,ozer2019investigation}. 
We find the sound velocity to be generally larger in GeO$_2$ by a factor from 1.22$\sim$1.38. 
The ratio is consistent with the inverse square root of the atomic mass of Ge and Sn ($(m_{\text{Ge}}/m_{\text{Sn}})^{-1/2}=1.28$).
The consistency is expected as the acoustic phonons are dominated by the heavy atoms, whose group velocity is inversely dependent on the square root of their atomic mass\cite{guo2021revisiting}. 
The higher speed of sound in GeO$_2$ contributes to its higher thermal conductivity, as confirmed by our subsequent calculations in section ~\ref{sec:therm_cond}. 


\begin{table*}
    \caption{Sound velocities (in km/s) for the LA ($v_{LA}$) and TA ($v_{TA}$) modes along the $\perp c$  and $\parallel c$ directions, and directional averages in SnO$_2$ and GeO$_2$. Our calculated values agree well with theoretical study in Ref.\onlinecite{bushick2020}, as well as experimental characterizations listed in the Table from Ref.\onlinecite{ozer2019investigation} for SnO$_2$ (directionally averaged) and Ref.\onlinecite{wang1973} for GeO$_2$.  }
    \begin{tabularx}{\textwidth}{l|LLLLLLLLLL}
        Sound velocity &$v_{\text{TA}}, \perp c$&$v_{\text{TA}}, \parallel c$&$v_{\text{LA}}, \perp c$&$v_{\text{LA}}, \parallel c$& $v^{ave}_{\text{TA}}$&$v^{ave}_{\text{LA}}$&$v_{\text{TA}}, \perp c$ (Expt.)&$v_{\text{TA}}, \parallel c$ (Expt.)&$v_{\text{LA}}, \perp c$ (Expt.)&$v_{\text{LA}}, \parallel c$ (Expt.) \\ \hline
        SnO$_2$ &4.41&3.50&5.78&7.75&4.10&6.44&\multicolumn{2}{c}{3.79}&\multicolumn{2}{c}{6.75} \\ 
        GeO$_2$&5.65 &4.86&7.09&9.48&5.39&7.89&5.072&6.415&7.328&9.770\\ \hline
    \end{tabularx}
    \label{tab:v_sound}
\end{table*}

\begin{figure}
    \centering
    \includegraphics[width=\columnwidth]{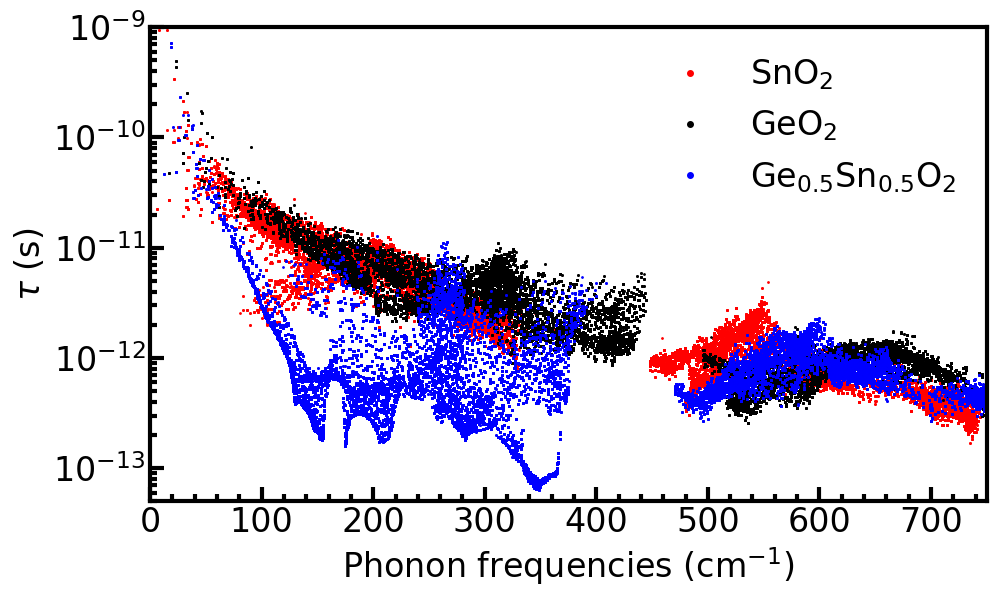}
    \caption{Phonon lifetimes due to three-phonon scattering processes for SnO$_2$, GeO$_2$, and Ge$_{0.5}$Sn$_{0.5}$O$_2$. Phonon lifetimes generally decrease with increasing phonon frequencies and are similar for the two binary compounds. A much steeper decrease is observed for the alloy, particularly for the low-frequency acoustic phonons.}
    \label{fig:tau}
\end{figure}

We further analyze the relaxation times of the phonons as a function of phonon frequencies, as shown in Figure~\ref{fig:tau}. 
For both SnO$_2$ and GeO$_2$, the relaxation times of the acoustic phonons decrease over three orders of magnitude from $10^{-9}$s to $10^{-12}$s with increasing phonon frequency from 0 to 400 cm$^{-1}$, while the relaxation times for the optical phonons are approximately on the order of $10^{-12}$s in general. 
Therefore, the low-frequency acoustic modes are expected to play a significant role in enabling high thermal conductivity for both binary compounds. 
However, a significant impact from alloying is observed in Ge$_x$Sn$_{1-x}$O$_2$, for which the phonon relaxation times display a much steeper decrease with increasing frequency. 
This effect is particularly pronounced in the frequency range below 400 cm$^{-1}$, where phonons in the alloy exhibit significantly lower relaxation times due to alloy disorder. 
The substantial decrease in phonon relaxation times for the alloy suggests a notably lower thermal conductivity, which we demonstrate in the following section.

\section{Thermal conductivity\label{sec:therm_cond}}
\subsection{Thermal conductivity with respect to temperature and alloy composition}

\begin{figure}
    \centering
    \includegraphics[width=\columnwidth]{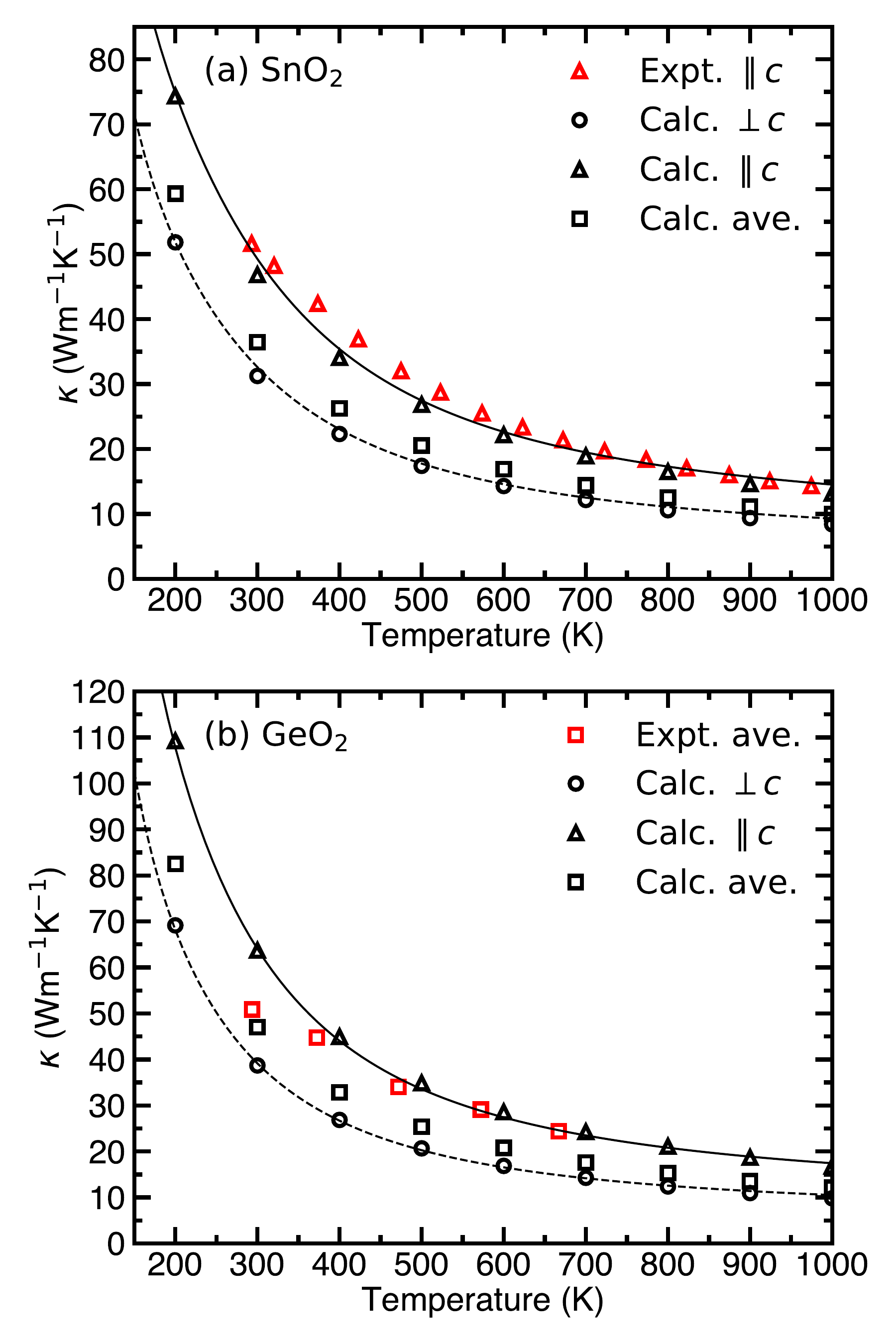}
    \caption{Temperature-dependent thermal conductivities of (a) SnO$_2$ and (b) GeO$_2$, reported both along the $\parallel c$ and $\perp c$ directions, as well as a directional averages for comparison to polycrystalline samples. 
    Fitted curves according a two-mode model [Eq.(\ref{eq:fit1})] and parameters from Tab.\ref{tab:fit} are shown as dashed and solid lines for $\perp c$ and $\parallel c$ directions.
    Good agreement is achieved compared to experimental measurements both for SnO$_2$\cite{galazka2020transparent}(single crystal, $\parallel c$) and GeO$_2$\cite{chae2020} (polycrystal).}
    \label{fig:therm_cond_end}
\end{figure}

The calculated temperature-dependent thermal conductivity of binary SnO$_2$ and GeO$_2$ (figure~\ref{fig:therm_cond_end}) is in good agreement with experimental measurements for both materials.\cite{galazka2020transparent,chae2020} For both materials, the dependence of their thermal conductivity can be described by the following equation: 
\begin{equation}\label{eq:fit1}
    \kappa(T)=\left[\frac{1}{\kappa_1}e^{-T_1/T}+\frac{1}{\kappa_2}e^{-T_2/T}\right]^{-1}
\end{equation}
In this equation, $\kappa_1$ and $\kappa_2$ are in units of thermal conductivities, and $T_1$, and $T_2$ are in units of temperatures. 
The model assumes a similar two-mode model illustrated in Ref.\onlinecite{bushick2020} for electron mobility with the two $\kappa$'s and $T$'s representing the contribution to thermal conductivity of the two different modes. 
The values of these parameters fitted from a full set of temperature- and composition-dependent model can be found in Table~\ref{tab:fit}. 
Figure~\ref{fig:therm_cond_end} shows that the fit well describes the temperature dependence of the thermal conductivity for both materials of the entire temperature range. 
GeO$_2$ shows a higher thermal conductivity for both modes. 
While the temperature of the high-energy modes are similar between the two materials, the temperature of the low-energy mode is higher in GeO$_2$. 
This trend is consistent with the difference between the light-massed cation, contributing mainly to the low energy mode, while the common oxygen anion contributes mainly to the high energy mode.

\begin{figure}
    \centering
    \includegraphics[width=0.9\columnwidth]{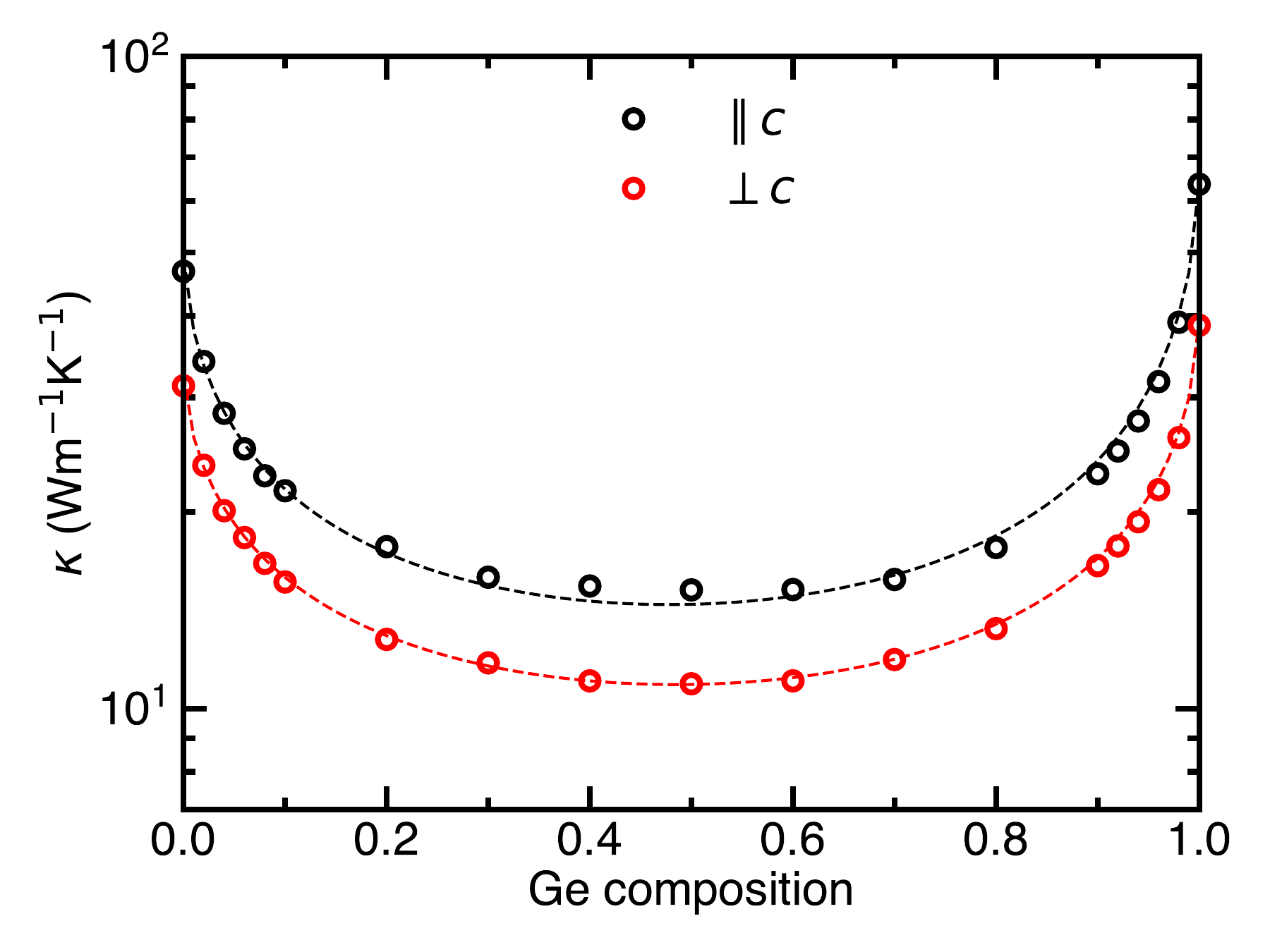}
    \caption{Calculated thermal conductivity of Ge$_x$Sn$_{1-x}$O$_2$ alloys at room temperature (300 K) as a function of alloy composition and crystallographic orientation as well as fitted curves (dashed) from the model described by Eq.(\ref{eq:fit2}), and parameters shown in Tab.\ref{tab:fit}. The thermal conductivity of the alloy is significantly reduced by alloy scattering, but even the lowest value remains comparable to that of Ga$_2$O$_3$ [11-15 Wm$^{-1}$K$^{-1}$ ($\perp c$), 15-20 Wm$^{-1}$K$^{-1}$ ($\parallel c$)]\cite{galazka2020transparent}. }
    \label{fig:300k}
\end{figure}

We next present the calculated thermal conductivities of the Ge$_{0.5}$Sn$_{0.5}$O$_2$ alloys at room temperature in figure~\ref{fig:300k}. The thermal conductivity of the alloy decreases significantly as the composition deviates from the pure compounds. At 10\% Ge or Sn content, the thermal conductivity drops to 43\% and 31\% of that of pure SnO$_2$ and GeO$_2$, respectively. The thermal conductivities across alloy compositions from $x=0.1$ to $x=0.9$ are reduced within a factor of 0.65 from the values at $x=0.1$ and $x=0.9$. The composition-dependent thermal conductivity at room temperature is described by the following equation:
\begin{equation}\label{eq:fit2}
    \kappa^{300K}(x)=\left[\frac{(1-x)^{\alpha}}{\kappa^{300K}_{\text{SnO}_2}}+\frac{x^{\alpha}}{\kappa^{300K}_{\text{GeO}_2}}+\frac{[x(1-x)]^{\alpha}}{\kappa'_{300K}}\right]^{-1}
\end{equation}
In this equation, $\kappa'_{300K}$ is a bowing parameter used to model the effect of alloy disorder\cite{fit1,fit2}, and $\alpha$ is a measure of the decay of the thermal conductivity close to the two binary compound. 
The fitted values of the parameters are $\alpha=0.668$ and $0.687$ for the $\perp c$ and $\parallel c$ direction, while the bowing at 300 K is obtained as $\kappa'_{300K}=7.05$ Wm$^{-1}$K$^{-1}$ for the $\perp c$ and $\kappa'_{300K}=8.34$ Wm$^{-1}$K$^{-1}$ for the $\parallel c$ direction. (See table \ref{tab:fit})
Comparing the result of the thermal conductivity of the Ge$_{0.5}$Sn$_{0.5}$O$_2$ alloy to $\beta$-Ga$_2$O$_3$, we find that both the in-plane ($\perp c$) and out-of-plane ($\parallel c$) thermal conductivity of the alloy are comparable to that of the pure Ga$_2$O$_3$ [11-15 Wm$^{-1}$K$^{-1}$ ($\perp c$), 15-20 Wm$^{-1}$K$^{-1}$ ($\parallel c$)]\cite{galazka2020transparent}. 
Our findings suggest that, despite the significant reduction in their thermal conductivity due to alloying, the heat dissipation in Ge$_{0.5}$Sn$_{0.5}$O$_2$-based devices is comparable to that of Ga$_2$O$_3$-based devices.

\begin{table*}
    \caption{Fitted values for the temperature- and alloy-composition-dependent thermal conductivity model of Equation~\ref{eq:fit3}. Parameters $\kappa_{1,\text{Sn}}, \kappa_{1,\text{Ge}}, \kappa_{2,\text{Sn}}, \kappa_{2,\text{Ge}}$, and $\kappa'$ are reported in units of thermal conductivity (Wm$^{-1}$K$^{-1}$), $T_{1,\text{Sn}}, T_{1,\text{Ge}}, T_{2,\text{Sn}}, T_{2,\text{Ge}}$, and $T'$ are in units of temperature (K), while $\alpha$ is a dimensionless parameter.
  The fitted values are also used to obtain describe temperature dependence for the end compounds [Eq.(~\ref{eq:fit1})] as well as composition dependence at 300 K [Eq.(\ref{eq:fit2})]. 
    }
    \begin{tabularx}{\textwidth}{l|LLLLLLLLLLLL}
         & $\alpha$&$\kappa_{1,\text{Sn}}$&$\kappa_{2,\text{Sn}}$&$T_{1,\text{Sn}}$&$T_{2,\text{Sn}}$&$\kappa_{1,\text{Ge}}$&$\kappa_{2,\text{Ge}}$&$T_{1,\text{Ge}}$&$T_{2,\text{Ge}}$&$\kappa'$&$T'$  \\ \hline
        $\perp c$ & 0.668 & 25.3 & 5.05 & 159 & 986 & 27.3 & 5.88 & 203 & 960& 5.59 & 69.6\\ 
        $\parallel c$&0.687 & 43.1 & 7.85 & 126 & 968 & 49.8 & 9.55 & 175 & 950 & 6.97 &53.9\\ \hline
    \end{tabularx}
    \label{tab:fit}
\end{table*}

\begin{figure}
    \centering
    \includegraphics[width=\columnwidth]{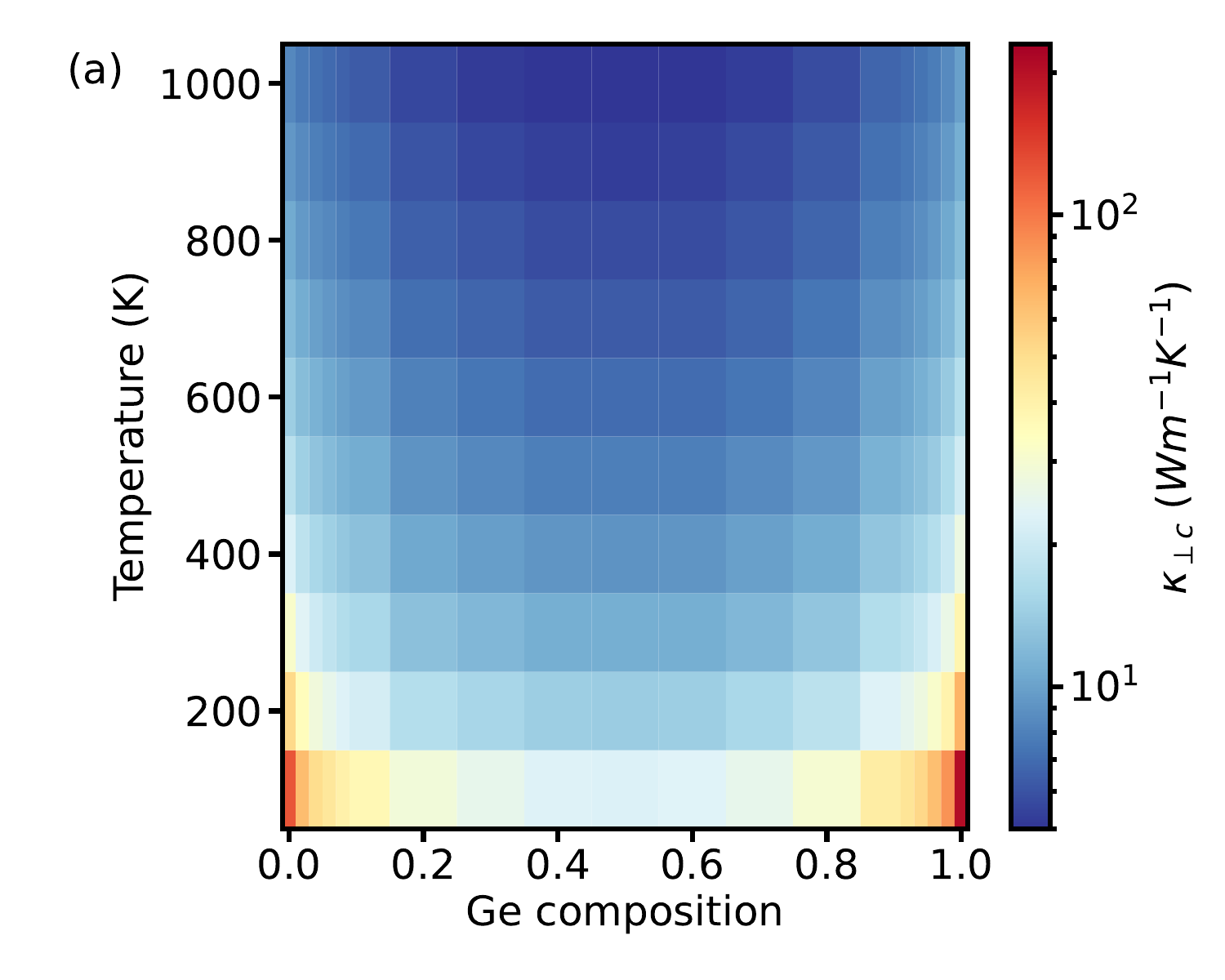}
    \includegraphics[width=\columnwidth]{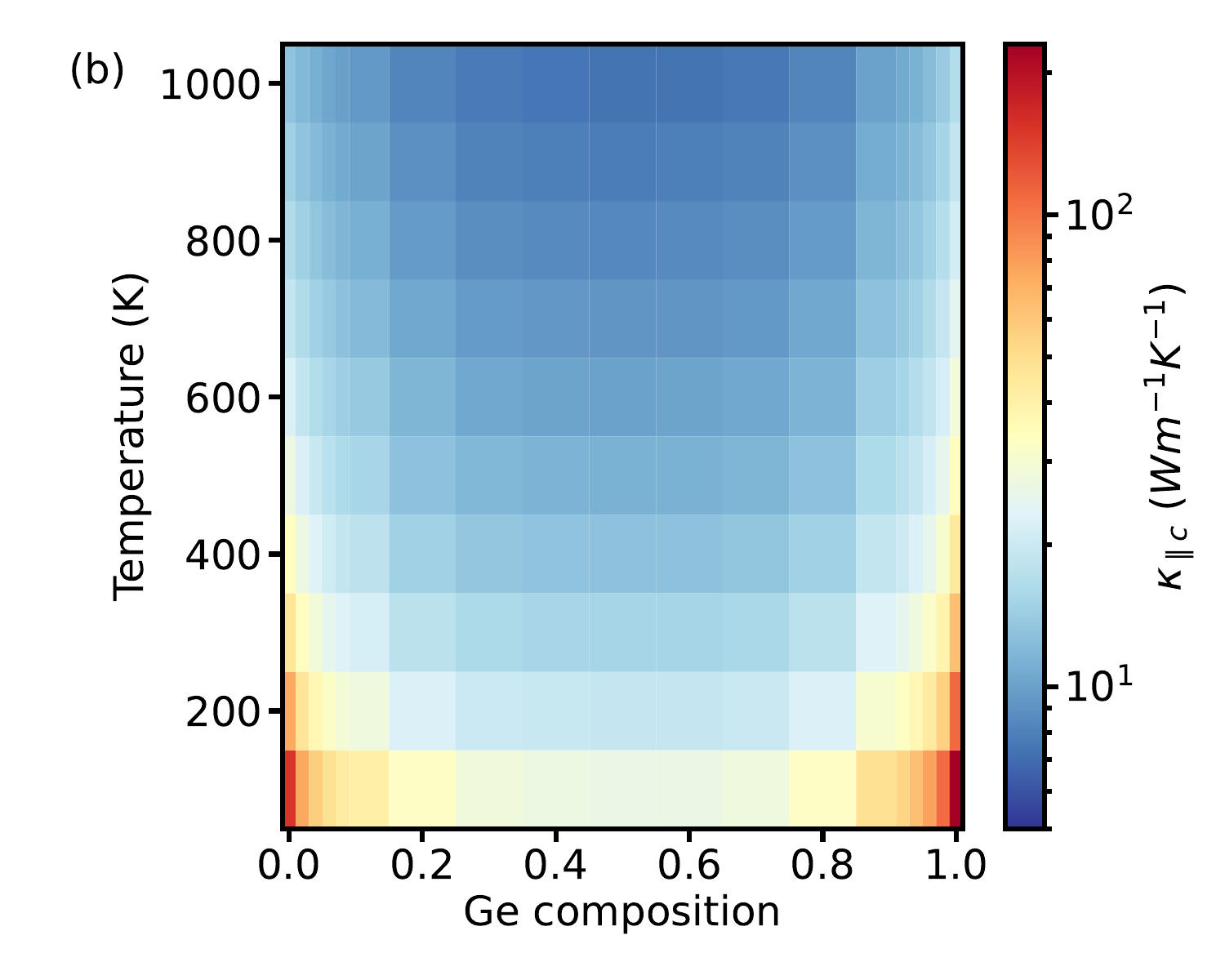}
    \caption{Thermal conductivity of SnO$_2$ and GeO$_2$ alloys as a function of alloy composition and temperature, along the (a) $\perp c$ and (b) $\parallel c$  direction. The significant reduction of thermal conductivity in the alloys suggest strong alloy scattering across the entire temperature range investigated. (See full tabulated data in supplemental information.)}
    \label{fig:therm_cond_alloy}
\end{figure}

\begin{figure}
    \centering
    \includegraphics[width=\columnwidth]{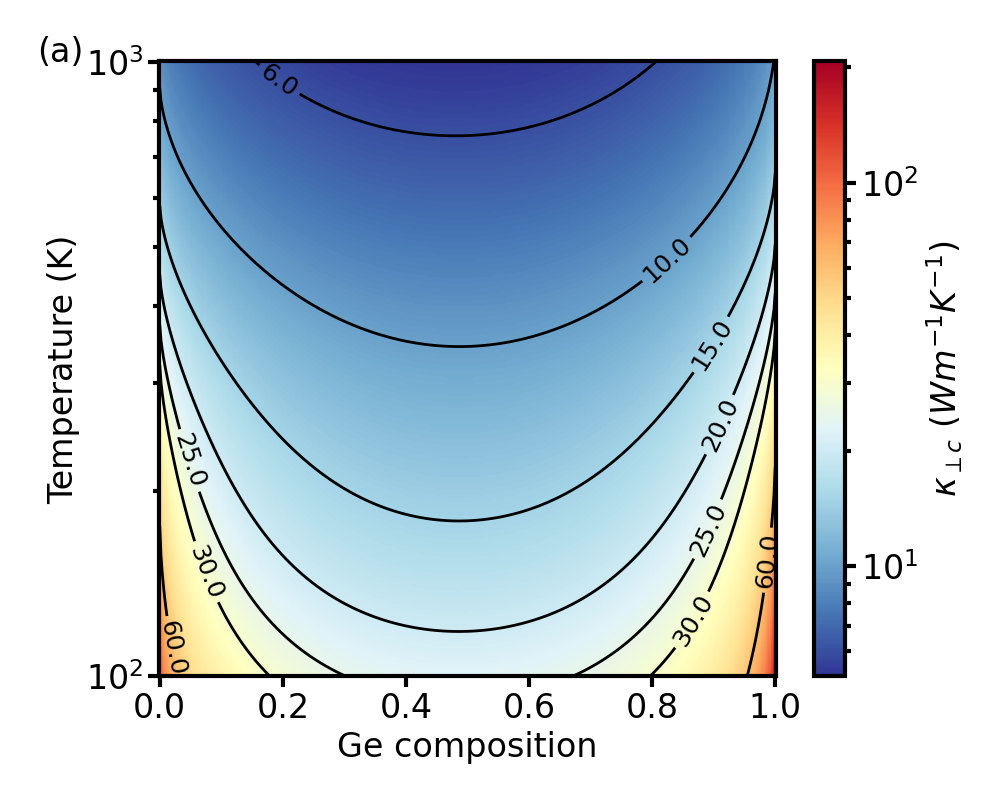}
    \includegraphics[width=\columnwidth]{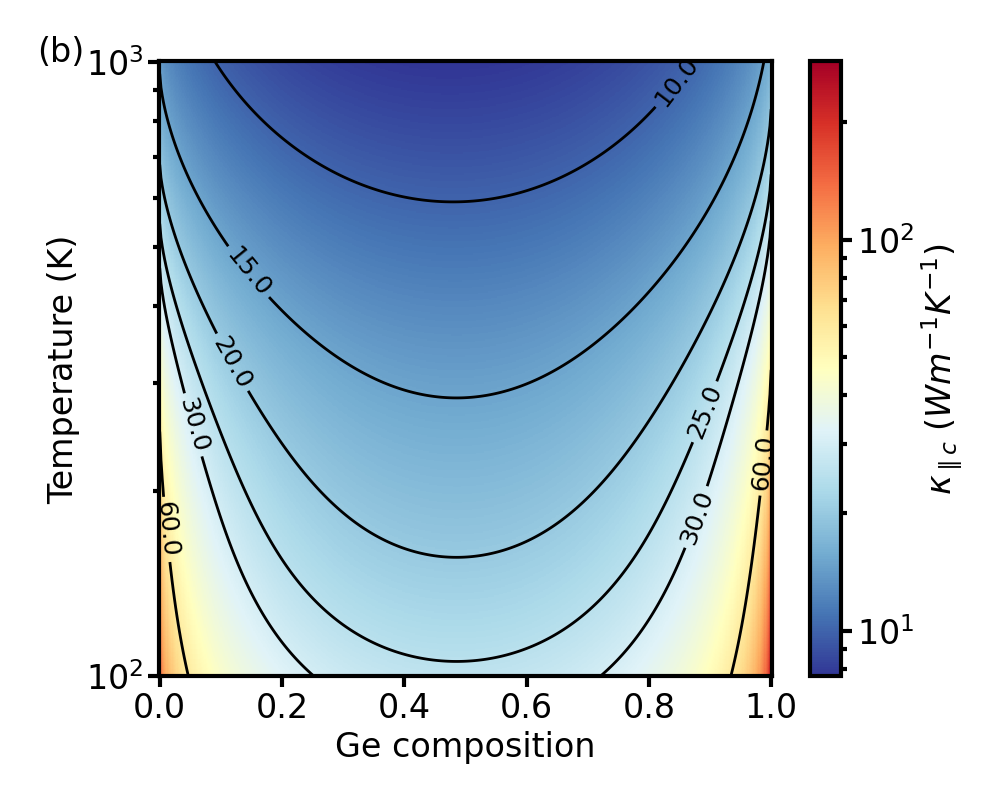}
    \caption{\label{fig:fit_all}Plot of the thermal conductivity as a function of Ge composition and temperature with Eq.(\ref{eq:fit3}) and the fitted parameters illustrated in Table~\ref{tab:fit}, as well as contour lines to demonstrate the values. 
    The maximum variation of the values from the calculated ones is 9.7\% and 10\% in the $\perp c$ and $\parallel c$ direction, respectively. 
    }
    \label{fig:isotope}
\end{figure}

Next, we present the temperature- and composition-dependent thermal conductivity of Ge$_{0.5}$Sn$_{0.5}$O$_2$ alloys in figure~\ref{fig:therm_cond_alloy}. (See tabulated data in the supplemental information) The thermal conductivity is fitted with the following equation, which combines the effects of temperature and composition dependence:
\begin{widetext}
\begin{equation}\label{eq:fit3}
    \kappa(x,T)=\left\{(1-x)^\alpha\left[\frac{1}{\kappa_{1,\text{Sn}}}e^{-T_{1,\text{Sn}}/T}+\frac{1}{\kappa_{2,\text{Sn}}}e^{-T_{2,\text{Sn}}/T}\right]+x^\alpha\left[\frac{1}{\kappa_{1,\text{Ge}}}e^{-T_{1,\text{Ge}}/T}+\frac{1}{\kappa_{2,\text{Ge}}}e^{-T_{2,\text{Ge}}/T}\right]+\frac{[x(1-x)]^\alpha}{\kappa'}e^{-T'/T}\right\}^{-1}
\end{equation}
\end{widetext}
In this equation, the parameters $\kappa_{1,\text{Sn}}, \kappa_{1,\text{Ge}}, \kappa_{2,\text{Sn}}, \kappa_{2,\text{Ge}}$, and $\kappa'$ are in units of thermal conductivity, $T_{1,\text{Sn}}, T_{1,\text{Ge}}, T_{2,\text{Sn}}, T_{2,\text{Ge}}$, and $T'$ are in units of temperature, while $\alpha$ is a dimensionless exponent. 
The values of the fitted parameters are listed in table~\ref{tab:fit}, and the fitted equations are plotted as a 2D heat map with isolines, shown in Fig.~\ref{fig:fit_all}.
In the entire temperature and composition range, the maximum difference between the fitted values and  calculated values is  9.7\% and 10\% in the $\perp c$ and $\parallel c$ direction, respectively. 
In the case of a nearly equimolar alloy ($x=0.5$), the significant reduction of the thermal conductivity due to strong alloy disorder renders a weak dependence when temperature is further considered. 
For example, the thermal conductivity along $\perp$c for the equimolar alloy reduces from 19 Wm$^{-1}$K$^{-1}$ to 5 Wm$^{-1}$K$^{-1}$ from 100 K to 1000 K, while for pure GeO$_2$, a reduction over an order of magnitude is seen from 155 Wm$^{-1}$K$^{-1}$ to 9.6 Wm$^{-1}$K$^{-1}$. 
This trend of strong alloy disorder surpassing the temperature dependence is also observed in III-V alloys such as AlGaN\cite{Huang_2022}.

\subsection{Impact of phonon mean-free-path and isotope scattering}

\begin{figure}
    \centering
    \includegraphics[width=\columnwidth]{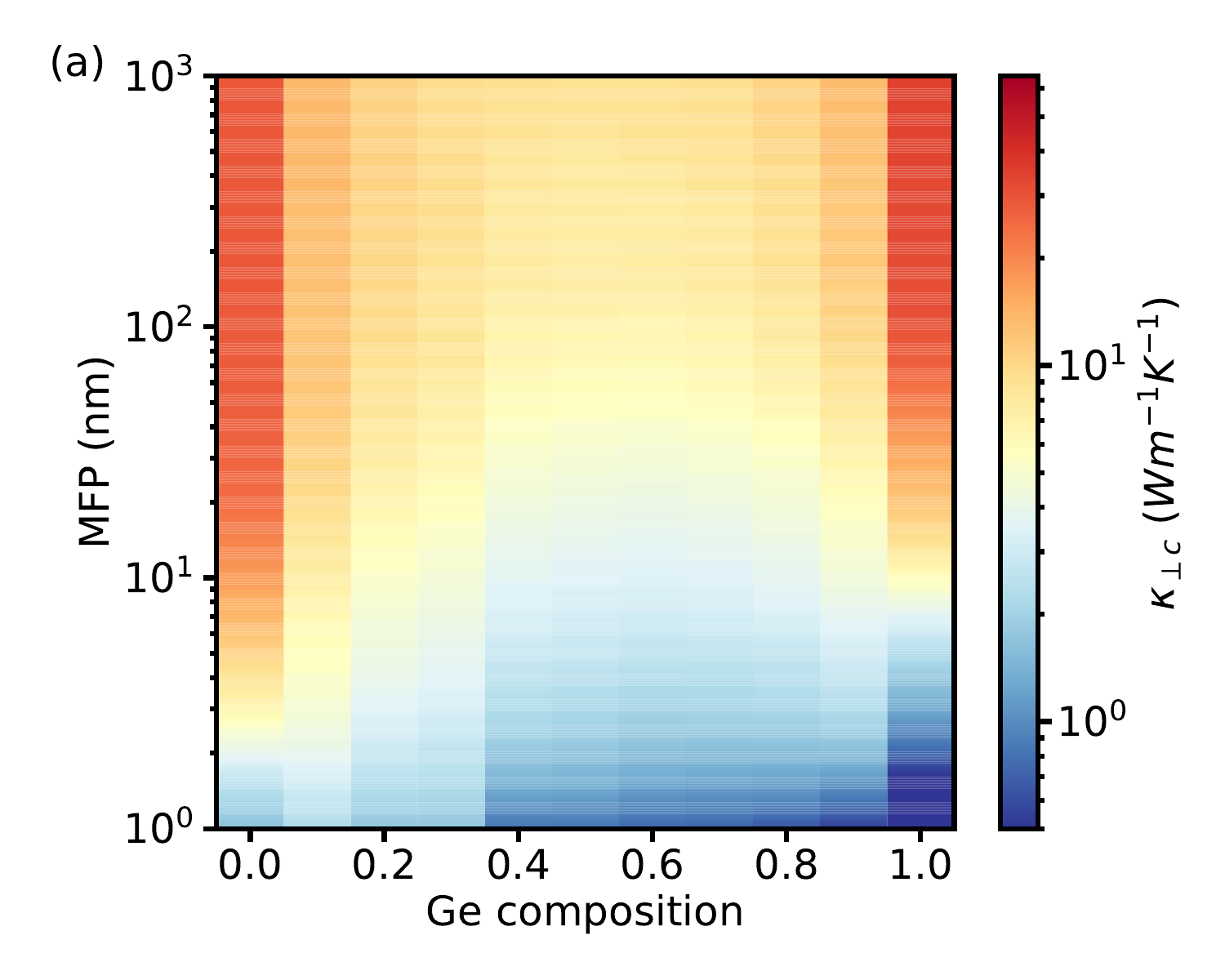}
    \includegraphics[width=\columnwidth]{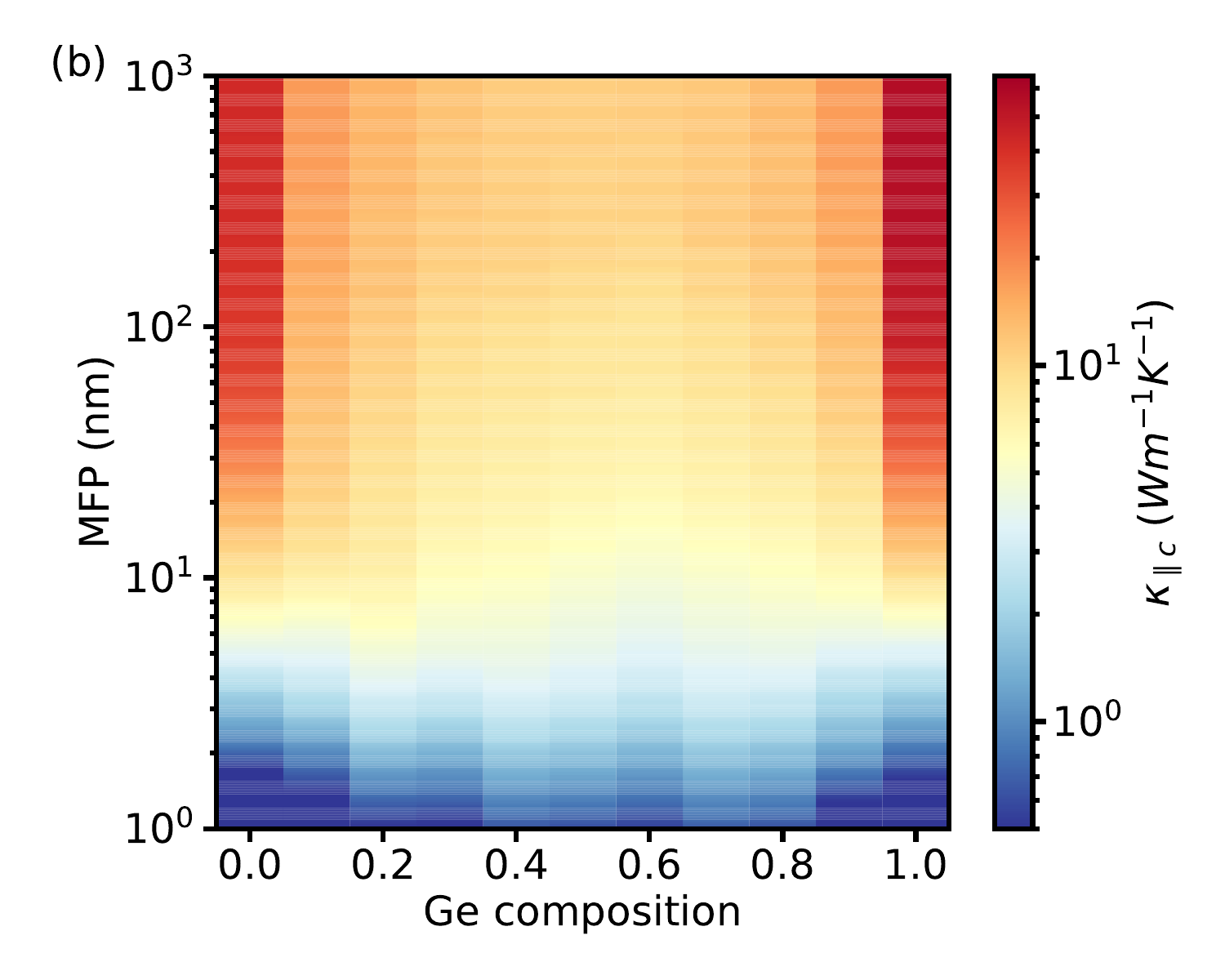}
    \caption{Cumulative thermal conductivities of Ge$_{0.5}$Sn$_{0.5}$O$_2$ at 300 K as a function of Ge content and the mean free path of the phonon projected to the respective directions for the (a) $\perp c$ and (b) $\parallel c$ direction. The thermal conductivity is strongly affected by the mean free path in the region of 10 - 100 nm. (See tabulated data in supplemental information.)}
    \label{fig:mfp}
\end{figure}

We further analyze the dependence of the thermal conductivity on the mean free path of the phonons to provide insight on the effect of scattering by grain boundaries in polycrystalline samples. In figure~\ref{fig:mfp}, we show the cumulative thermal conductivity at room temperature (300 K) evaluated within the relaxation time approximation, including only those phonons that have a mean free path up to the specified value. 
This analysis offers an evaluation of the fundamental upper limit of the lattice thermal conductivity for a given average grain size. 
We find a sharp increase of the phonon-limited thermal conductivity as the phonon mean-free-path increases from 10 nm to 100 nm, which is consistent with a previous theoretical study for GeO$_2$\cite{chae2020}. 
In general, it is expected that high-quality crystals result in reduced scattering by grain boundaries, therefore benefiting thermal conductivity. 
In Table~\ref{tab:mfp}, we list the values of the phonon mean free path required to achieve 80\% of the ideal thermal conductivity for SnO$_2$, GeO$_2$, and the equimolar alloy in the $\perp c$ and $\parallel c$ direction. 
Our calculations quantify the grain sizes required to achieve desired theoretical thermal conductivity limits for efficient thermal management.

\begin{table}
    \centering
        \caption{Phonon mean free path (in nm) required for the thermal conductivity to reach 80\% of the ideal value for SnO$_2$, GeO$_2$, and Ge$_{0.5}$Sn$_{0.5}$O$_2$ at 300 K. For all cases, 80\% of the ideal value of thermal conductivity is achieved for grain sizes above 100 nm. }
    \begin{tabular}{c|ccc}
    Compound&SnO$_2$&GeO$_2$&Ge$_{0.5}$Sn$_{0.5}$O$_2$  \\ \hline
         $\perp c$& 431 & 102 &437 \\
          $\parallel c$& 112 & 132 & 362 \\ \hline
    \end{tabular}
    \label{tab:mfp}
\end{table}

\begin{figure}
    \centering
    \includegraphics[width=0.49\columnwidth]{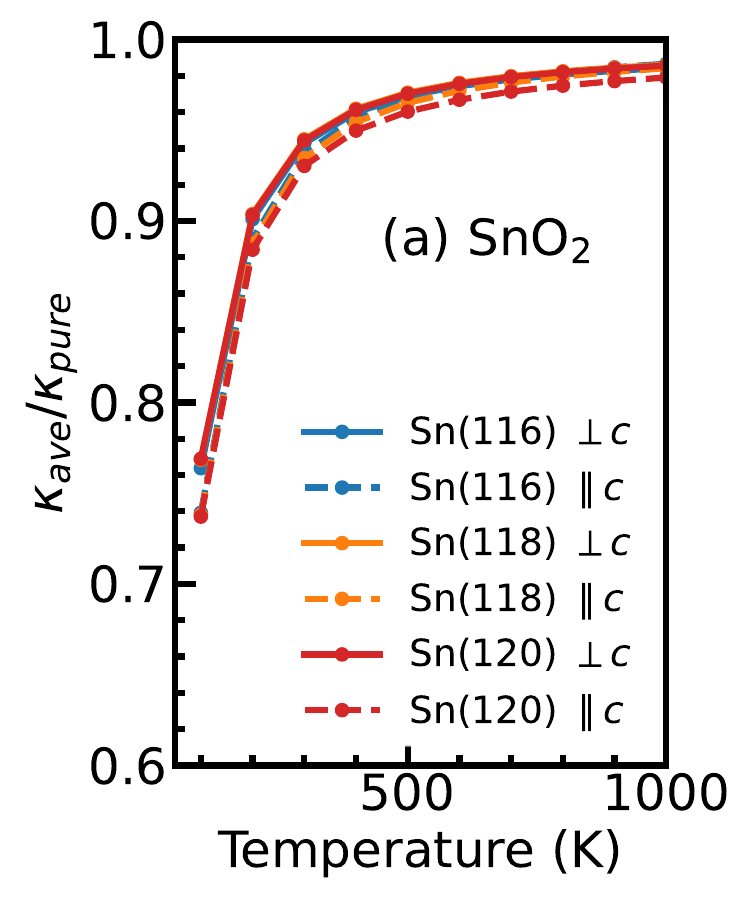}
    \includegraphics[width=0.49\columnwidth]{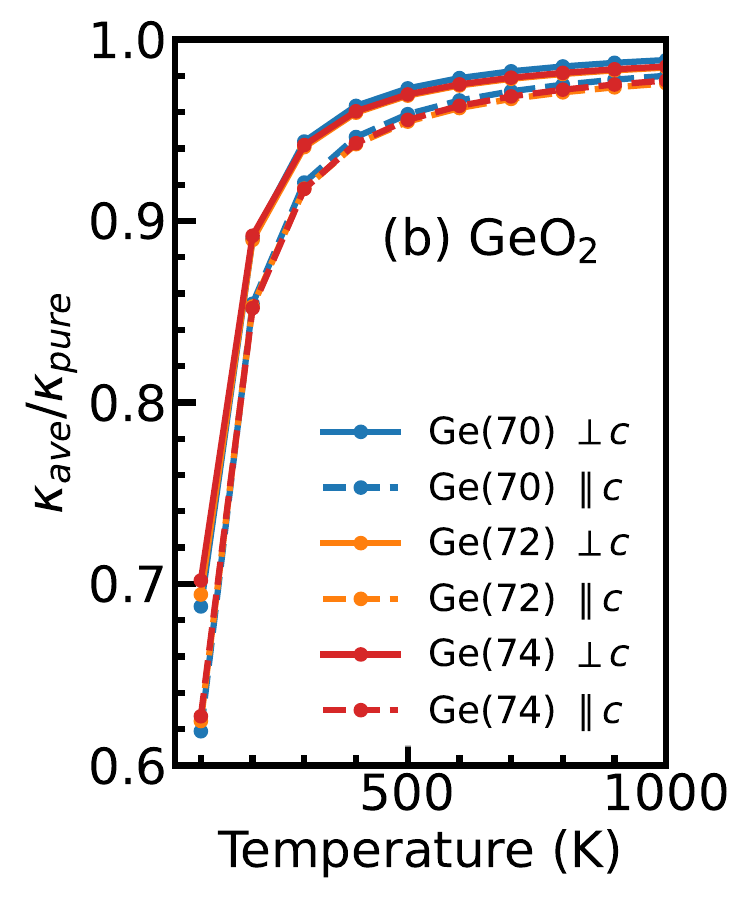}
    \caption{Ratio of the thermal conductivity of (a) SnO$_2$ and (b) GeO$_2$ evaluated with the average atomic mass considering isotope mass disorder ($\kappa_{ave}$) to that with a pure isotope ($\kappa_{pure}$).
    The three most abundant isotopes are considered for each material. 
    For both materials, isotope scattering has a stronger impact on the thermal conductivity at low temperatures and along the $\parallel c$ direction.
    }
    \label{fig:isotope}
\end{figure}

Due to the different masses of the various isotopes of a given atomic species, isotope mass disorder is expected to affect the thermal conductivity. We quantify the effects of isotope scatterings in SnO$_2$ and GeO$_2$, and we uncover a stronger effect of isotope scattering in GeO$_2$. 
In figure~\ref{fig:isotope}, we show the effect of isotope scattering on thermal conductivity by plotting the ratio of thermal conductivity evaluated with the average isotope mass including mass disorder to that of the isotope pure material considering the three most abundant species, as a function of temperature for both SnO$_2$ and GeO$_2$. 
This ratio provides a measure of how the scattering induced by isotope mass disorder affects thermal transport across different temperatures. SnO$_2$ shows a more modest reduction across all temperatures compared to GeO$_2$. 
At low temperatures (e.g. 100 K), isotope scattering significantly reduces the thermal conductivity of GeO$_2$ by 30\% in the $\perp c$ and 38\% in the $\parallel c$ direction, compared to a more moderate reduction in SnO$_2$ of 23\% in the $\perp c$ and 26\% in the $\parallel c$.
At higher temperature such as room temperature (300 K), the impact of isotope scattering are less significant: Isotope mass disorder is responsible for a reduction of thermal conductivity by 6\% in the $\perp c$ and 8\% in the $\parallel c$ direction for GeO$_2$, and by 5\% in the $\perp c$ and 6\% in the $\parallel c$ direction for SnO$_2$. 
As temperature further increases, the relative impact of isotope mass disorder on phonon transports decreases as the ratio gradually approaches unity. These findings provide insight into the fundamental limitations on thermal conductivity in GeO$_2$-based materials, particularly at lower temperatures.

\section{Conclusion}

In this study, we provide a comprehensive analysis of the thermal conductivity of GeO$_2$, SnO$_2$, and their alloys utilizing first-principles calculations. Our findings reveal that alloying significantly reduces thermal conductivity due to enhanced phonon scattering, particularly for the acoustic phonon modes. Despite this reduction, the thermal conductivity of the alloys remains comparable with $\beta$-Ga$_2$O$_3$. 
We also produced a complete analytical model that fits the first-principles thermal conductivity data of Ge$_{0.5}$Sn$_{0.5}$O$_2$ alloys across various compositions and temperatures.
We found a two-mode model to well describe the temperature- and composition-dependent thermal conductivity of the alloy, offering valuable predictive power for optimizing alloy design.

Our analysis further shows that polycrystalline samples with grain sizes less than 100 nm reduce the phonon mean-free-path and therefore reduce the thermal conductivity. 
The influence of Ge and Sn isotopes scattering is found to be considerable at low temperatures, and becomes less dominant as temperature increases. 
We find isotope mass disorder to reduce the thermal conductivity of the binaries by 5-8\% at room temperature, providing insights into the fundamental limitations of the thermal conductivity in GeO$_2$-based materials.

Our findings provide a fundamental understanding of the alloy-, phonon-, and isotope-disorder-limited thermal conductivity in Ge$_{0.5}$Sn$_{0.5}$O$_2$ alloys as a function of composition and temperature, and offer guidance for the codesign of these materials for high-power electronic applications, where both tunable electronic properties and efficient heat dissipation are essential.

\acknowledgments
This material is based upon work supported by the National Science Foundation under grant no. 2328701 and is supported in part by funds from federal agency and industry partners as specified in the Future of Semiconductors (FuSe) program. Computational resources were provided by the National Energy Research Scientific Computing Center, which is supported by the Office of Science of the U.S. Department of Energy under Contract
No. DE-AC02-05CH11231.

\clearpage
\bibliography{aipsamp}

\end{document}